\documentclass[10pt,aps,prb,twocolumn,superscriptaddress]{revtex4-1}

\usepackage{natbib}
\usepackage[dvips]{graphicx}
\usepackage{bm}
\usepackage{amsmath, amssymb}

\newcommand{\BFA}{BaFe$_{2}$As$_{2}$}
\newcommand{\SFA}{SrFe$_{2}$As$_{2}$}
\newcommand{\CFA}{CaFe$_{2}$As$_{2}$}
\newcommand{\TN}{$T_{\mathrm{N}}$}
\newcommand{\Ts}{$T_{\mathrm{s}}$}
\newcommand{\Tc}{$T_{\mathrm{c}}$}
\newcommand{\cm}{cm$^{-1}$}
\newcommand{\Neff}{$N_{\mathrm{eff}}$}
\newcommand{\sa}{$\sigma_a$}
\newcommand{\sib}{$\sigma_b$}
\newcommand{\saw}{$\sigma_a(\omega)$}
\newcommand{\sbw}{$\sigma_b(\omega)$}

\begin{document}



\title{Unprecedented anisotropic metallic state in \BFA{} revealed by optical spectroscopy}





\author{M.~Nakajima}
\email[]{nakajima@lyra.phys.s.u-tokyo.ac.jp}
\affiliation{Department of Physics, University of Tokyo, Tokyo 113-0033, Japan}
\affiliation{National Institute of Advanced Industrial Science and Technology, Tsukuba 305-8568, Japan}
\affiliation{JST, Transformative Research-Project on Iron Pnictides (TRIP), Tokyo 102-0075, Japan}
\author{T.~Liang}
\affiliation{Department of Physics, University of Tokyo, Tokyo 113-0033, Japan}
\affiliation{National Institute of Advanced Industrial Science and Technology, Tsukuba 305-8568, Japan}
\affiliation{JST, Transformative Research-Project on Iron Pnictides (TRIP), Tokyo 102-0075, Japan}
\author{S.~Ishida}
\affiliation{Department of Physics, University of Tokyo, Tokyo 113-0033, Japan}
\affiliation{National Institute of Advanced Industrial Science and Technology, Tsukuba 305-8568, Japan}
\affiliation{JST, Transformative Research-Project on Iron Pnictides (TRIP), Tokyo 102-0075, Japan}
\author{Y.~Tomioka}
\affiliation{National Institute of Advanced Industrial Science and Technology, Tsukuba 305-8568, Japan}
\affiliation{JST, Transformative Research-Project on Iron Pnictides (TRIP), Tokyo 102-0075, Japan}
\author{K.~Kihou}
\affiliation{National Institute of Advanced Industrial Science and Technology, Tsukuba 305-8568, Japan}
\affiliation{JST, Transformative Research-Project on Iron Pnictides (TRIP), Tokyo 102-0075, Japan}
\author{C.~H.~Lee}
\affiliation{National Institute of Advanced Industrial Science and Technology, Tsukuba 305-8568, Japan}
\affiliation{JST, Transformative Research-Project on Iron Pnictides (TRIP), Tokyo 102-0075, Japan}
\author{A.~Iyo}
\affiliation{National Institute of Advanced Industrial Science and Technology, Tsukuba 305-8568, Japan}
\affiliation{JST, Transformative Research-Project on Iron Pnictides (TRIP), Tokyo 102-0075, Japan}
\author{H.~Eisaki}
\affiliation{National Institute of Advanced Industrial Science and Technology, Tsukuba 305-8568, Japan}
\affiliation{JST, Transformative Research-Project on Iron Pnictides (TRIP), Tokyo 102-0075, Japan}
\author{T.~Kakeshita}
\affiliation{Department of Physics, University of Tokyo, Tokyo 113-0033, Japan}
\affiliation{JST, Transformative Research-Project on Iron Pnictides (TRIP), Tokyo 102-0075, Japan}
\author{T.~Ito}
\affiliation{National Institute of Advanced Industrial Science and Technology, Tsukuba 305-8568, Japan}
\affiliation{JST, Transformative Research-Project on Iron Pnictides (TRIP), Tokyo 102-0075, Japan}
\author{S.~Uchida}
\affiliation{Department of Physics, University of Tokyo, Tokyo 113-0033, Japan}
\affiliation{JST, Transformative Research-Project on Iron Pnictides (TRIP), Tokyo 102-0075, Japan}


\begin{abstract}
An ordered phase showing remarkable electronic anisotropy in proximity to the superconducting phase is now a hot issue in the field of high-transition-temperature superconductivity. As in the case of copper oxides, superconductivity in iron arsenides competes or coexists with such an ordered phase. Undoped and underdoped iron arsenides have a magnetostructural ordered phase exhibiting stripe-like antiferromagnetic spin order accompanied by an orthorhombic lattice distortion; both the spin order and lattice distortion break the tetragonal symmetry of crystals of these compounds. In this ordered state, anisotropy of in-plane electrical resistivity is anomalous and difficult to attribute simply to the spin order and/or the lattice distortion. Here, we present the anisotropic optical spectra measured on detwinned \BFA{} crystals with light polarization parallel to the Fe planes. Pronounced anisotropy is observed in the spectra, persisting up to an unexpectedly high photon energy of about 2 eV. Such anisotropy arises from an anisotropic energy gap opening below and slightly above the onset of the order. Detailed analysis of the optical spectra reveals an unprecedented electronic state in the ordered phase.
\end{abstract}

\maketitle

\section{Introduction}

High-transition-temperature (high-\Tc{}) superconductivity realized in both copper oxides and iron arsenides shares common features, namely, the superconducting phase is in close proximity to a symmetry-breaking phase and these phases coexist under certain circumstances, but apparently compete with each other. The close proximity suggests that our understanding of high-\Tc{} superconductivity will greatly improve once the nature of this proximate phase is revealed. The parent compounds of iron-arsenide superconductors, with \BFA{} as a representative example, are unique metals which undergo a tetragonal-to-orthorhombic structural phase transition at temperature \Ts{} with a shorter $b$ axis and a longer $a$ axis in the orthorhombic phase always accompanied by antiferromagnetic (AF) spin order at temperature \TN{}. \TN{} is equal to \Ts{} in some compounds \cite{Huang2008,Zhao2008,Goldman2008} and slightly lower than \Ts{} in others. \cite{Cruz2008} \BFA{} exhibits stripe-like AF order in which Fe spins align antiferromagnetically in the $a$-axis direction in the Fe plane and ferromagnetically in the $b$-axis direction. Anisotropic electronic properties have been experimentally examined by various methods, such as neutron scattering, \cite{Zhao2009} scanning tunneling microscopy (STM), \cite{Chuang2010} and angle-resolved photoemission spectroscopy (ARPES). \cite{Shimojima2010,Yi2010} These experiments suggest strong anisotropy of spin excitation and of the shape of Fermi surfaces. However, most of the experiments were performed on twinned crystals with randomly oriented domains, which inhibit the observation of genuine anisotropy.

Recently, anisotropic resistivity has been measured on detwinned crystals. \cite{Tanatar2010,Chu2010} The anisotropy of resistivity is quite anomalous in that the resistivity along the spin-ferromagnetic (FM) direction with a shorter $b$ axis is higher than that along the spin-AF direction with a longer $a$ axis. This is quite counterintuitive, since normally electrons would hop more easily between atoms along the shorter $b$ axis and along the direction of FM spin alignment; this is known as the spin-double-exchange mechanism, which gives rise to colossal magnetoresistance in doped manganites. Therefore, nontrivial mechanism to produce such anisotropy should be operative. A fundamental question is which degrees of freedom, lattice, electron spin, charge, or orbital, play a primary role in the formation of this anisotropic ordered state.

The anisotropic electronic state or broken rotational symmetry, often called nematicity, is currently one of the hot issues in research of high-\Tc{} superconducting copper oxides. \cite{Fradkin2010,Daou2010} It is related to the nature of a mysterious pseudogap phase competing and/or coexisting with the $d$-wave superconducting phase. A recent STM study revealed that the breaking of rotational symmetry originates from electronic differences between the two oxygen sites (O$_x$ and O$_y$) within each CuO$_2$ unit cell, which should be equivalent in the fourfold symmetric tetragonal lattice. \cite{Lawler2010} For iron pnictides, it has been suggested that the lattice symmetry breaking from tetragonal to orthorhombic structure is driven either by anisotropic spin fluctuations (or spin nematic state) \cite{Fang2008,Xu2008,Mazin2009,Fernandes2010} or by orbital ordering associated with a distinction and repopulation of the Fe 3$d_{xz}$ and 3$d_{yz}$ electron orbitals, which are degenerate in energy in the tetragonal Fe sublattice. \cite{Lee2009,Kruger2009,Yin2010a,Daghofer2010,Lv2010,Chen2010,Valenzuela2010} Experimentally, ARPES study indicates that the orbital character near the Fermi energy is strongly polarized; \cite{Shimojima2010} however, the net change in orbital occupancy is likely small. \cite{Yi2010}

Here, to elucidate the intrinsic anisotropic electronic properties and their origins of the parent compounds of iron-pnictide superconductors, we study the optical conductivity of a single-domain \BFA{} crystal with the electric field of polarized light along the $a$- and $b$-axis directions. The optical spectroscopy is a bulk-sensitive and energy-resolved probe and is suitable for investigating of lattice (phonon) and electronic excitations along the energy axis. Optical conductivity ($\sigma(\omega)$) is the absorptive current response to a time-varying external electric field of frequency $\omega$; hence, it is an extension of the dc ($\omega$=0) conductivity to finite energies. A similar optical measurement on a detwinned \BFA{} crystal has been done also by Dusza \textit{et al.}, \cite{Dusza2010} but the spectral feature in the low-energy data is not clear enough to inquire the connection to the origin of the anisotropy in resistivity. The present work provides comprehensive optical spectra of \BFA{} covering wide energy and temperature ranges.

\section{Experimental}

Single crystals of \BFA{} were grown by the self-flux method as described elsewhere. \cite{Nakajima2010} We chose crystals with area and thickness suitable for optical measurement under uniaxial pressure. The crystals were cut in a rectangular shape, the sides of which were parallel to the tetragonal [110] direction. Typical crystal dimensions were 3.5 mm $\times$ 3.5 mm $\times$ 0.5 mm. The crystals were sealed with BaAs in an evacuated quartz tube and annealed at 800 $^\circ$C for 5 days. After annealing, the residual resistivity ratio (RRR) increased from 3 to $\sim$ 30. The annealed crystal exhibited magnetostructural phase transition at \Ts{}=143 K, among the highest reported so far. This is probably because annealing removes defects at Ba or As sites, lattice dislocations, or both, and suppresses carrier scattering.

\begin{figure}
\includegraphics[width=83mm,clip]{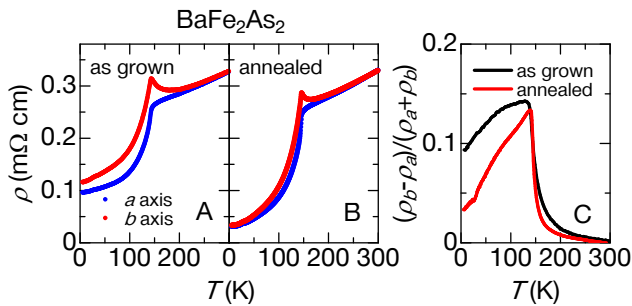}%
\caption{Temperature dependence of the in-plane resistivity of detwinned \BFA{}, $\rho_a$ (blue) and $\rho_b$ (red). (A) For the as-grown crystal, anisotropy starts at $\sim$ 220 K and persists down to the lowest measurement temperature. (B) Annealing decreases residual resistivity and increases magnetostructural transition temperature $T_{\mathrm{s}}$ (at the peak in $\rho_b$). (C) Annealing markedly decreases the overall anisotropy, both the magnitude and temperature ranges.}
\end{figure}

To remove twinned domains, we applied compressive pressure (approximately 50 MPa) in one direction. The crystalline $b$ axis in the orthorhombic phase aligns along the pressure direction. Resistivity was measured using the Montgomery method, \cite{Montgomery1971} which enables us to measure resistivity along the $a$ and $b$ axes simultaneously. The temperature dependence and magnitude of the anisotropy of resistivity in the as-grown crystal (Fig.\ 1A) are similar to previous reports. \cite{Tanatar2010,Chu2010} The optical reflectivity $R(\omega)$ was measured with incident light almost normal to the $ab$ plane in the frequency range of 50-40000 \cm{} under uniaxial pressure for the $a$- and $b$-axis polarizations of the orthorhombic structure at various temperatures using a Fourier transform infrared spectrometer (Bruker IFS113v) and a grating monochromator (JASCO CT-25C). The optical conductivity $\sigma(\omega)$ was derived from the Kramers-Kronig (K-K) transformation of $R(\omega)$. The Drude-Lorentz formula was used for the low-energy extrapolation in order to connect smoothly to the spectrum in the measured region and to fit the measured resistivity at $\omega$=0.

\section{Results}

Figure 1A shows the temperature ($T$) dependence of the resistivity of as-grown \BFA{} along the $a$ and $b$ axes ($\rho_a$ and $\rho_b$, respectively). The measurements were conducted by applying uniaxial pressure of about 50 MPa to make the crystal almost twin-free. For the as-grown crystal, anisotropy appears at temperatures about 80 K higher than \Ts{} (=138 K) and persists down to the lowest measurement temperature. After annealing, the resistivity jumps up in the $b$ direction and drops in the $a$ direction at \Ts{} (=143 K), as shown in Fig.\ 1B. The anisotropy of resistivity appears only at temperatures 30-40 K above \Ts{}, and the resistivity becomes nearly isotropic at low temperatures. In Fig.\ 1C, note that the anisotropic resistivity ratio $(\rho_b-\rho_a)/(\rho_a+\rho_b)$, which may serve as a nematic order parameter, reaches a maximum slightly below \Ts{} and decreases as temperature is lowered. From the sharper transition in the annealed crystal, we speculate that it would be of first-order if the compound becomes much cleaner (a discontinuous first-order transition is more clearly seen in \SFA{} and \CFA{}). \cite{Krellner2008,Ni2008}

\begin{figure}
\includegraphics[width=83mm,clip]{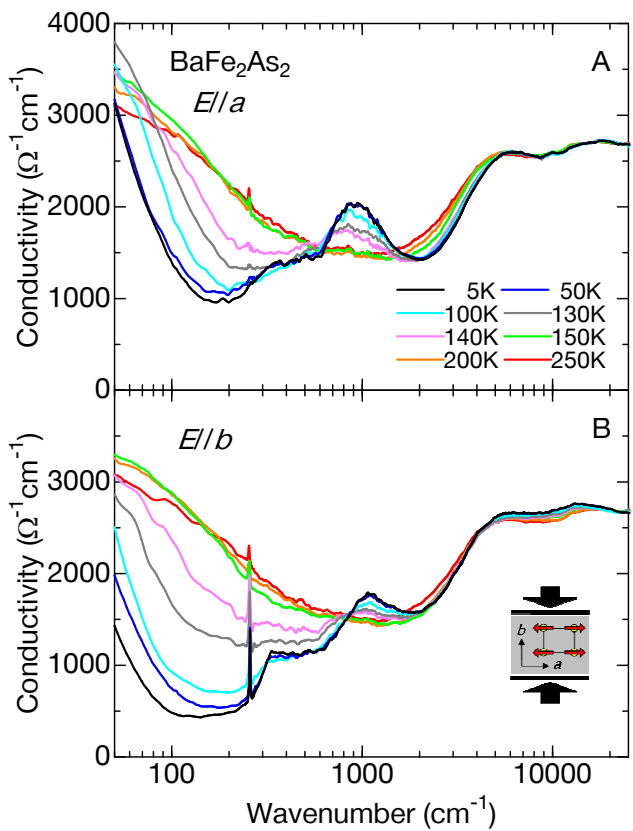}%
\caption{Temperature evolution of the optical conductivity spectrum of detwinned \BFA{} for polarization parallel to the $a$ and $b$ axes. The crystal is twin-free as long as compressive pressure is applied in one direction. The crystalline $a$ and $b$ axes, as well as the spin alignment with respect to the pressure direction, are shown in the inset. Rapid suppression of low-energy conductivity is seen in both spectra below 140 K, just below \Ts{}=143 K. Spectra at temperatures well above \Ts{} show weak $T$ dependence showing no incipient gap feature. A precursory gap is observed only for the $b$-axis spectrum at $T$=150 K. Singular features (a spike at 257 \cm{} and a cusp at 334 \cm{}) appear only in \sbw{}.}
\end{figure}

Although the anisotropy of resistivity is strongly diminished at low temperatures, it is found that the optical spectra show robust anisotropy. Figures 2A and 2B show the temperature evolution of the optical conductivity spectra of detwinned \BFA{} (after annealing) over a wide energy range for the $a$- and $b$-axis polarizations (\saw{} and \sbw{}), respectively. Both spectra are identical within the present experimental precision at temperatures above 200 K in stark contrast to the result taken by Dusza \textit{et al}. \cite{Dusza2010} Note that the measurement was carried out under uniaxial pressure even in the tetragonal phase above \Ts{}, evidencing that the applied pressure hardly affects the charge dynamics. A small difference is discernible between \sa{} and \sib{} only at $T$=150 K in the region at approximately 400 \cm{} (see Appendix B) where the \sib{} spectrum (in green) shows a slight decrease relative to the spectrum at 200 K (in orange). However, the \sa{} spectra at the two temperatures shows no significant change. As analyzed and discussed below, the decrease in \sib{} indicates the opening of a precursory gap associated with a fluctuating (or short-range) order.

Upon cooling to less than \Ts{}=143 K, a rapid reduction in conductivity is observed below $\sim$ 600 \cm{} in \sa{} and below $\sim$ 850 \cm{} in \sib{} reflecting a sudden onset of the long-range order at \Ts{}. The low-energy conductivity is more strongly suppressed in the $b$ direction. This suppression is partly due to the rapid narrowing of a conductivity peak at $\omega$=0 (a Drude peak), but mainly due to the opening of gaps. The opening of gaps is also evident from the appearance of a bump in \sa{} (a cusp in \sib{}) at $\sim$ 340 \cm{}, as well as by the growth of a peak centered at 950 \cm{} in \sa{} (1050 \cm{} in \sib{}), which is formed by accumulating the low-energy conductivity that is reduced due to the gap. From a comparison between the missing spectral weight (SW or \Neff{}($\omega$), the energy integral of the reduced conductivity with respect to that at $T$=200 K) in the low-energy gap region and the increased SW in the peak region, it is found that almost all the reduced conductivity is accumulated to form the peak in \saw{} (Appendix C). On the other hand, the missing SW in \sib{} is transferred not only to the peak at 1050 \cm{} but also to a much higher-energy region, as mentioned below. As a result, the anisotropy of conductivity persists up to 17000 \cm{} ($\sim$ 2 eV), as shown in the spectra at $T$=5 K in Fig.\ 3. Below 1350 \cm{}, \sib{} is lower than \sa{}, whereas above 1350 \cm{} the anisotropy is reversed. Note that the gap feature appearing at $\sim$ 340 \cm{} in both \sa{} and \sib{} plays a minor role in this SW transfer.

\begin{figure}
\includegraphics[width=83mm,clip]{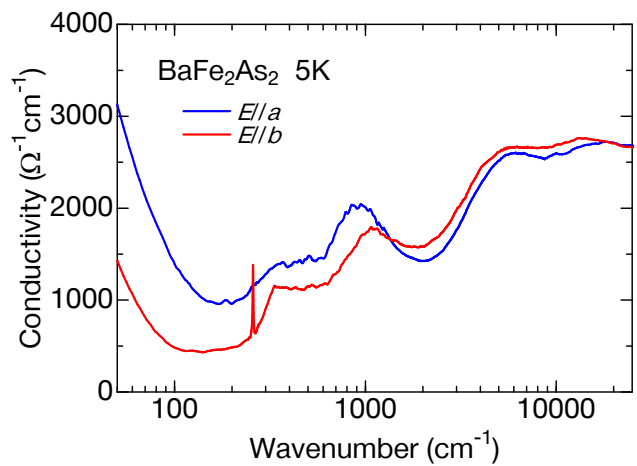}%
\caption{Comparison between the $a$- and $b$-axis optical conductivity spectra at 5 K above 50 \cm{}. Below 1350 \cm{}, \sib{} is lower than \sa{}, whereas above 1350 \cm{} the anisotropy is reversed. The anisotropy persists up to an energy as high as 17000 \cm{} ($\sim$ 2 eV). Note that the anisotropy decreases and approaches 1 toward $\omega$=0.}
\end{figure}

In the region from 1500 to 5000 \cm{}, a reduction in conductivity is observed for \saw{} with a decrease in temperature from 250 K. The increased conductivity in the higher-energy region between 5000 and 9000 \cm{} exactly compensates for the reduced spectral weight, indicative of the opening of another gap or a blueshift of the energy of the interband transition between the relevant energy bands. Independent of the SW transfer below 1500 \cm{}, the transfer in the region from 1500 to 5000 \cm{} starts at $T$=300 K or higher, progresses steadily down to 140 K, and then slows down.

The high-energy \sib{} spectrum also shows a similar temperature dependence down to $T$=200 K. However, below 200 K, the decrease in conductivity between 1500 and 5000 \cm{} is modest, but its increase above 5000 \cm{} (up to 17000 \cm{}) is appreciable, particularly below 150 K. Clearly, the reduced SW below 850 \cm{} is transferred to this energy region, accounting for the 1050-\cm{} peak which is not so developed in comparison with the 950-\cm{} peak in \saw{}. This result indicates that the magnetostructural order affects the $b$-axis conductivity over a wide energy range up to 17000 \cm{} ($\sim$ 2 eV); however, its effect extends only up to 1500 \cm{} in \saw{}. Note that these spectral changes, and the resulting anisotropy on such a large energy scale, are not compatible with an ordinary spin-density-wave (SDW) scenario based on the Fermi-surface nesting, but rather are indicative of strong reconstruction of electronic energy bands upon entering the ordered state. The effect of orthorhombic lattice distortion below \Ts{} (and under small applied uniaxial pressure) cannot modify the electronic structure over such a large energy scale. Also note that the fluctuation of this order, observed as an incipient gap feature in the $b$-axis optical spectrum, fades away at $T$=200 K in contrast to the spin excitation spectrum in which the anisotropic spin fluctuation of the stripe AF order persists well above 200 K. \cite{Diallo2010} Thus, the anisotropic gap behavior might not be directly linked to the spin order, but rather related to the orbital degree of freedom, which would drive structural deformation as suggested by the recent ARPES measurement on detwinned \BFA{}. \cite{Yi2010}

\section{Discussion}

\begin{figure}
\includegraphics[width=83mm,clip]{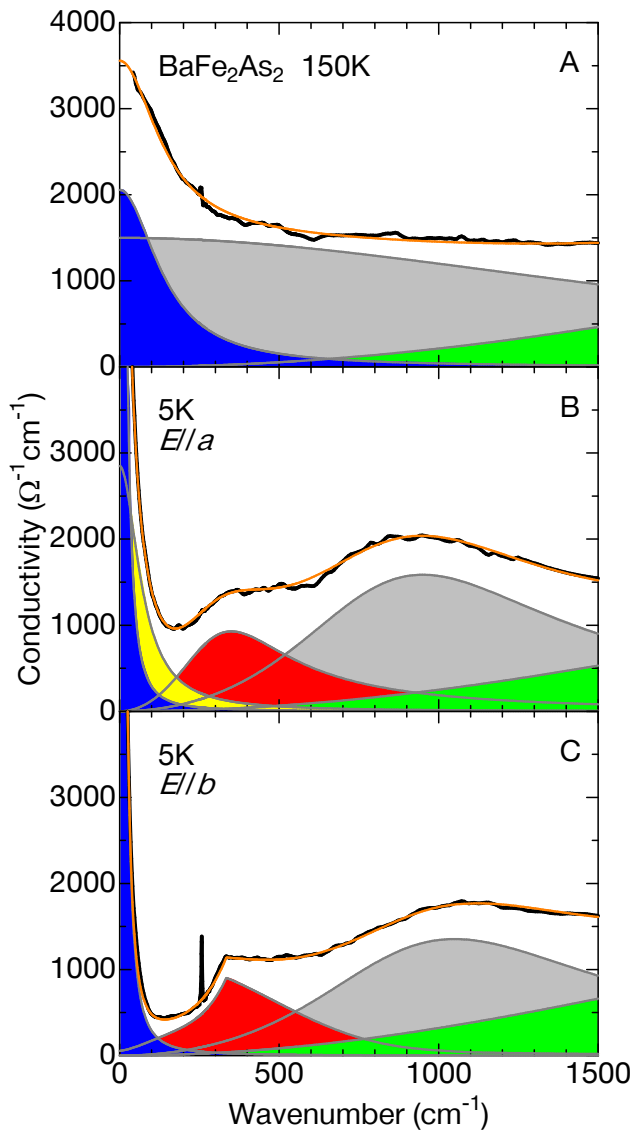}%
\caption{Decomposition of the low-energy conductivity spectrum below 1500 \cm{}. (A) The spectrum at 150 K in the tetragonal-paramagnetic phase is decomposed into a broad Drude component and a narrow Drude component, both contributing to the dc conductivity. (B, C) In the orthorhombic-stripe AF phase ($T$=5 K), the width of the narrow Drude component in A becomes extremely narrow and the peak height reaches 30000 $\Omega^{-1}$\cm{} in both spectra. A gap opens in the broad Drude component (in gray), but the gap in \sib{} is deeper than that in \sa{}. In the gap region a small Drude component (yellow) and an absorption band (red) remain in \sa{}, while in \sib{} only a singular band (red) remains and a spike appears.}
\end{figure}

The emergence of anisotropy and of certain spectral features observed in this experiment is explained by band structure calculations to some extent. \cite{Yin2010b,Sanna2010,Laad2010} Several calculations demonstrated that the stripe AF order, which breaks the fourfold tetragonal symmetry, induces the reconstruction of the electronic structure over a wide energy range (several tenths of an electronvolt) by lifting the degeneracy of Fe 3$d_{xz}$ and 3$d_{yz}$ orbitals. \cite{Lee2009,Kruger2009,Yin2010a,Daghofer2010,Lv2010,Chen2010,Valenzuela2010} As a consequence, the Fermi surface is composed of mainly 3$d_{xz}$ orbital, while the energy bands composed of 3$d_{yz}$ orbital shift away from the Fermi level, thus forming gaps for electronic excitations. The distinction between the 3$d_{xz}$ and 3$d_{yz}$ orbitals thus leads to a difference in optical response between the polarized light along the $a$ ($x$) and $b$ ($y$) axes. In fact, the calculation of optical conductivity by Yin, Haule, and Kotliar \cite{Yin2010b} using local density approximation (LDA) combined with dynamical mean-field theory (DMFT) is qualitatively in good agreement with the present experimental results, although the gap energy scales, particularly that in \sbw{}, are different. In addition, there are certain spectral features observed in the ordered state that are not reproduced by theoretical calculations, such as the two singular features, that is, a spike at 257 \cm{} and a cusp at 334 \cm{}, observed in \sbw{}.

Below, we analyze the experimental data by decomposing the spectrum into several components. Given that iron pnictides are multiband systems, this is a quite natural approach to analyzing the result.\cite{Nakajima2010,Wu2010} As shown in Fig.\ 4A, the low-energy spectra at $T$=150 K just above \Ts{}=143 K for both polarizations are nearly identical, and each spectrum is roughly decomposed into two Drude terms: one (in gray) is much broader than the other (in blue). With increasing temperature, the narrower Drude component broadens, whereas the broader Drude component shows no appreciable change. \cite{Nakajima2010}

In the ordered phase below \Ts{}, the spectrum in each direction is decomposed differently as in Figs.\ 4B and 4C for the data at $T$=5 K. It turns out that opening of gaps is observed in the broader Drude component in the spectrum above \Ts{}. Common to both spectra, the Drude component becomes extremely narrow and its peak height becomes extremely large (exceeding 30000 $\Omega^{-1}$\cm{} at $T$=5 K) corresponding to a very low resistivity in both directions at low temperatures. A small difference is the longer tail lasting above 100 \cm{} of the $\omega$=0 peak in \saw{}, which requires an additional Drude component that probably originates from an ungapped portion of the Fermi surface after reconstruction. This component with a peak height of $\sim$ 3000 $\Omega^{-1}$\cm{}, one order of magnitude lower than that of the main Drude component, may explain the slightly lower resistivity in the $a$ direction.

One insight gained from this analysis is that the conductivity anisotropy arises not from the Drude component, \cite{Yi2010} but rather from higher-energy components which show a deeper gap in the $b$ direction than in the $a$ direction. The underlying cause of such anisotropy is most likely the distinction between Fe 3$d_{xz}$ and 3$d_{yz}$ orbitals (orbital polarization), as suggested by theory \cite{Lee2009,Kruger2009,Yin2010a,Daghofer2010,Lv2010,Chen2010,Valenzuela2010} and ARPES experiment. \cite{Shimojima2010,Yi2010}

Another insight is that the Drude component is almost isotropic and is extremely narrow at low temperatures ($\sim$ 5 \cm{} at $T$=5 K). This extreme narrowness indicates that carrier scattering from impurities and/or lattice defects is severely suppressed. Theoretical calculations \cite{Yin2010b,Ran2009,Morinari2010} and a recent ARPES experiment \cite{Richard2010} show the existence of small Fermi surface pockets originating from the energy bands with Dirac-cone-shaped dispersions. Theoretically, the unique topological nature of such a band structure inhibits the backscattering of electrons from impurities. \cite{Morinari2010} This makes the electron mobility very high; thus, the pronounced Drude peak in the optical conductivity spectrum is likely associated with Dirac-cone Fermi surface pockets.

The Drude term rapidly broadens and hence the peak height rapidly decreases with increasing temperature and with increasing Co dopant concentration. This is also the case with the unannealed crystals. Accordingly, contribution of higher-energy components to the dc conductivity becomes appreciable, which explains the enhanced anisotropy of resistivity as temperature approaches \Ts{} and as doped Co density increases \cite{Chu2010}, as also observed in the unannealed samples.

The conductivity component which appears as a bump in \saw{} is expressed by a Lorentzian function that peaks at 350 \cm{} and tails up to 1500 \cm{} (shown in red). The corresponding component in \sbw{} is quite singular, showing a spike-shaped peak at 257 \cm{} and a cusp at 334 \cm{}. These features are also observed in \sbw{} for detwinned \SFA{}, albeit somewhat smeared, and are not predicted by theoretical calculations. As shown in Fig.\ 4C, after the decomposition of \sbw{}, a singular absorption band emerges with a small width ranging from near zero energy to approximately 800 \cm{} ($\sim$ 0.1 eV). The energy ($\omega$) dependence of this band is asymmetric, showing $-\sqrt{\omega_{\mathrm{s}}-\omega}$ singularity on the lower-energy side of the cusp energy ($\omega_{\mathrm{s}}$) and slowly decreasing above $\omega_{\mathrm{s}}$. \cite{Suzuura1996} The origin of this singular band is not yet clear. Thus far, no excitation that gives rise to such a singular band has been predicted in \BFA{} or other pnictide systems.

The spike at 257 \cm{} is observed only in \sbw{}, but is hardly seen in \saw{} below \Ts{}. A peak at nearly the same energy is observed in the spectrum of \BFA{} in the tetragonal phase above \Ts{} and also in the spectra of other iron-arsenide compounds. \cite{Akrap2009,Dong2010} This peak is assigned to an optical phonon mode involving the displacement of both Fe and As atoms. Akrap \textit{et al.} have investigated the $T$ dependence of this peak in detail for a twinned crystal of \BFA{}. \cite{Akrap2009} They reported anomalous $T$ dependence of the energy ($\omega_0$) and intensity (oscillator strength) of this peak, both of which exhibit a discontinuous shift at \Ts{}. Notably, the peak's intensity is enhanced below \Ts{} by a factor of 2. In the present experiment on a single-domain crystal, it turns out that the intensity of the 257-\cm{} spike in \sbw{} increases by a factor of 3-4 in comparison with that above \Ts{}, whereas its intensity in \saw{} decreases to an indiscernible level.

The 257-\cm{} peak is extremely narrow, $\sim$ 3 \cm{} or less, even though it is superposed on the continuum (red component) showing the cusp. This implies that the coupling with this excitation band is weak. The sharpness appears to be linked to the gap opening in another component (in gray). Thus, it may be natural to suppose that the corresponding mode is overdamped in the $a$ direction, owing to the strong coupling with the overlapping component (shown in red), which likely has different orbital character from the component in \sbw{}. The theoretical calculations show that \saw{} comes primarily from the 3$d_{xz}$ (and 3$d_{xy}$) orbital, and \sbw{} is associated with the 3$d_{yz}$ (and 3$d_{xy}$) orbital. \cite{Yin2010b} However, the distinction between the 3$d_{xz}$ and 3$d_{yz}$ orbitals alone has difficulty in explaining the considerable increase in the spike's intensity in one direction and its decrease in the other direction. An alternative scenario is that a new phonon mode appears in the ordered state which is located at the Brillouin zone edges of the tetragonal phase and is folded back to the zone center of the orthorhombic phase. The abrupt shift in $\omega_0$ across \Ts{}, as reported by Akrap \textit{et al.}, appears to support this scenario, but it is difficult to ascertain why the original phonon mode at $\omega_0$ loses its intensity in one direction below \Ts{}. The change in the intensity of the optical phonon mode appears to require an additional polarization field induced by this lattice vibration in the ordered state. The induced field would almost cancel out the dipole field associated with the optical phonon mode in the $a$ direction, while it would reinforce the dipole field in the $b$ direction. The origin of the induced polarization is not yet clear, but it should be generated by an electronic excitation process, likely related to the orbital degree of freedom.


The present results on the anisotropy of optical conductivity give deeper insight into the electronic state of the parent compound of iron-pnictide superconductors, \BFA{}. Low-energy conductivity is dominated by a nearly isotropic Drude component which rapidly narrows and grows in a peak height with decreasing temperature across \Ts{}, consistent with the reduced anisotropy of resistivity at low temperatures. Appreciable conductivity anisotropy appears in the higher-energy region and persists up to $\sim$ 2 eV. This anisotropy originates from the anisotropic gap opening in the ``incoherent'' component of the spectrum in the high-temperature paramagnetic-tetragonal phase associated with the onset of the long-range magnetostructural order. The anisotropy of resistivity becomes visible when the Drude peak height decreases with increasing temperature and probably increasing Co dopant concentration. \cite{Nakajima2010} The most striking anisotropy is observed for a possible optical phonon mode below \Ts{} which increases its intensity in \sbw{} but nearly vanishes in \saw{}. The present experiment also reveals a peculiar narrow absorption band with a singular $\omega$ dependence present only in the $b$-axis direction. Its origin is not yet known but is suggestive of a hidden singularity due to the electronic excitation in this compound. The results indicate that a novel electronic state is formed in the ordered state of \BFA{} in which the orbital degree of freedom may play a more important role than so far anticipated via unique coupling with charge, spin, and phonon.




\begin{figure}
\centering
\includegraphics[width=80mm,clip]{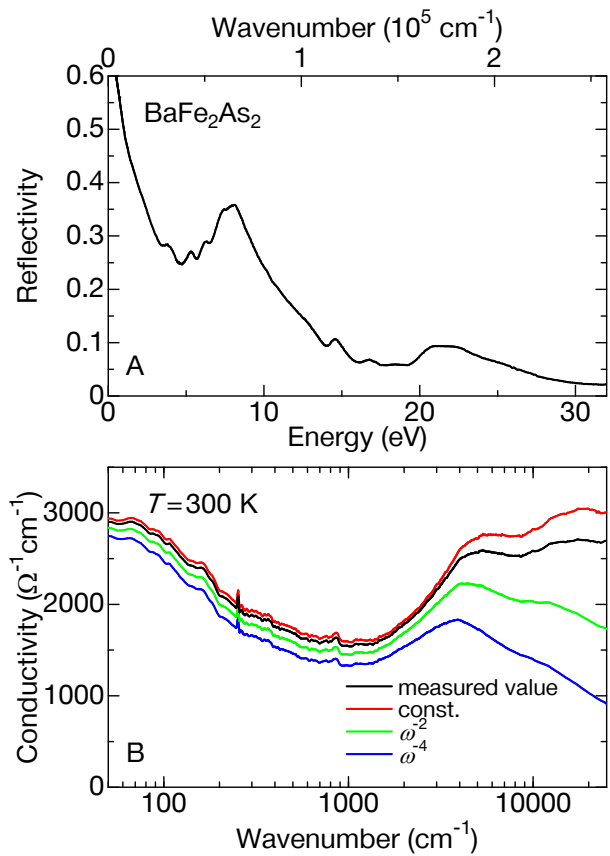}%
\caption{(A) Reflectivity spectrum of \BFA{} measured at room temperature and for the in-plane polarization in the energy range up to 32 eV. (B) Kramers-Kronig transformed optical conductivity spectra for various high-energy extrapolations.}
\end{figure}

\begin{acknowledgments}
M.N. and S.I. thank the Japan Society for the Promotion of Science (JSPS) for financial support. The authors thank T. Tohyama, H. Kontani, N. Nagaosa, and T. Timusk for helpful discussions. This work was supported by the Japan-China-Korea A3 Foresight Program from JSPS, and a Grant-in-Aid for Scientific Research from JSPS and the Ministry of Education, Culture, Sports, Science, and Technology, Japan.
\end{acknowledgments}


\appendix

\section{Reflectivity spectrum in the ultraviolet region}

\begin{figure}
\centering
\includegraphics[width=80mm,clip]{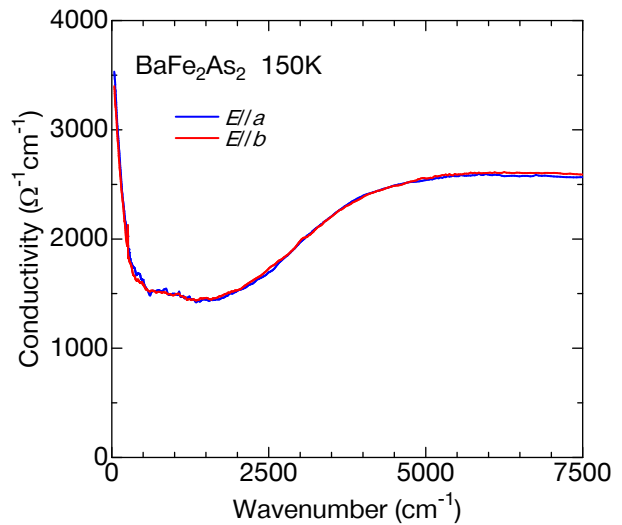}%
\caption{Optical conductivity spectrum of detwinned \BFA{} for the polarizations of the $a$ and $b$ axis at 150 K. Although a small difference appears in the low-energy region ($\sim$ 400 \cm{}), both of the $a$- and $b$-axis spectra are nearly identical.}
\end{figure}

\begin{figure}
\centering
\includegraphics[width=80mm,clip]{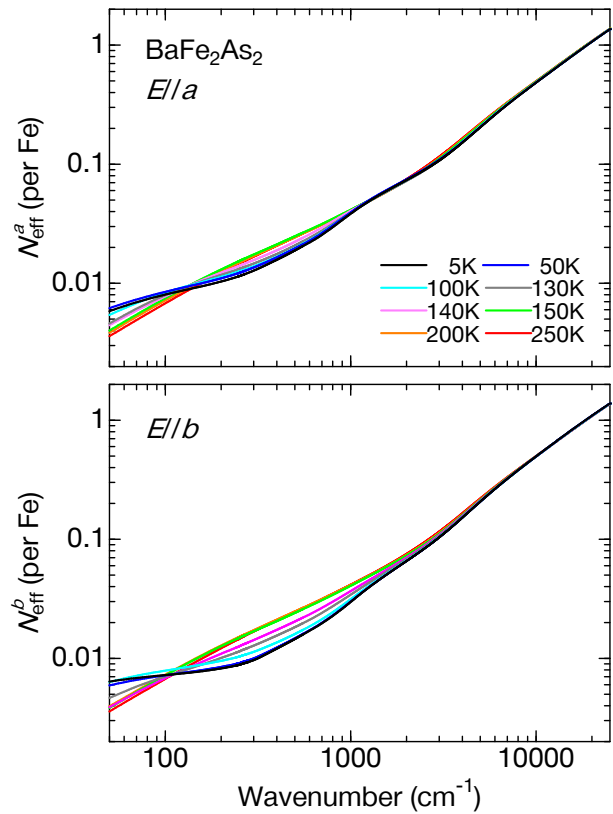}%
\caption{Integrals of conductivity (\saw{} and \sbw{} shown in Fig.\ 2) up to the energy $\omega$, \Neff{}($\omega$) for the $a$- and $b$-axis polarizations.}
\end{figure}

Reflectivity measurement in the vacuum-ultraviolet energy region above 4 eV ($\sim$ 32000 \cm{}) was performed at room temperature using a Seya-Namioka-type grating with synchrotron radiation as the light source at the UVSOR facility of the Institute for Molecular Science. This is necessary to obtain reliable optical conductivity via the K-K transformation of the reflectivity spectrum for multi-band/orbital systems such as the present compound or cuprates, \cite{Tajima1989} since a number of intraband and interband excitations overlap in the same energy region and extend up to 10 eV or higher. The spectrum up to 32 eV is shown in Fig.\ 5A. The reflectivity shows good coincidence with the result measured in the laboratory. As the measured optical reflectivity shows no detectable anisotropy or temperature dependence above $\sim$ 2 eV, this high-energy spectrum can be used to obtain optical conductivity at lower temperatures.

If we use conventional extrapolation schemes for the K-K transformation, the obtained conductivity appreciably deviates from the present result, particularly in the region higher than 2000 \cm{}. Figure 5B shows the conductivity spectra obtained by various extrapolations above 4.5 eV.

\section{Optical conductivity spectrum at 150 K}

For comparison with the result reported by Dusza \textit{et al.}, \cite{Dusza2010} our data of \saw{} and \sbw{} at $T$=150 K are displayed in Fig.\ 6 in linear scale. As described in the text, no significant difference is seen between the $a$- and $b$-axis spectra above $T$=200 K in the present work. Even at 150 K just above \Ts{}, the anisotropy is very small in contradiction to the data shown in Fig.\ 2 in Ref.\ \onlinecite{Dusza2010}.

\section{Effective-carrier-number analysis}

Based on the conductivity data, we calculated the spectral weight in terms of the effective carrier number per Fe site;
\begin{equation*}
N_{\mathrm{eff}} (\omega) = \frac{2 m_0 V}{\pi e^2} \int^{\omega}_{0} \sigma(\omega ^\prime) \mathrm{d} \omega ^\prime,
\end{equation*}
where $m_0$ is the free electron mass and $V$ the cell volume containing one Fe atom. \Neff{}$(\omega)$ is proportional to the number of carriers involved in the optical excitation up to a certain energy $\omega$. It provides information on the energy at which the specific excitation terminates and on the energy scale of the spectral weight transfer due to gap opening. The results are shown in Fig.\ 7. The plateau at approximately 100 \cm{} at low temperatures in $N_{\mathrm{eff}}^b$ indicates that a Drude component terminates at this energy and that this component is well separated from gap or interband excitations in the higher-energy region. For the $a$-axis polarization, \Neff{} curves merge at approximately 1500 \cm{}, evidencing that the reduced conductivity in the low-energy region due to gap opening is transferred to the higher-energy region up to 1500 \cm{}. On the other hand, \Neff{} curves for the $b$ axis merge at a much higher energy, indicating that the spectral weight transfer takes place over a very large energy scale (see Fig.\ 2B).

\end{document}